\documentclass[10pt,a4paper]{article}

\usepackage{indentfirst}			
 \usepackage{enumerate}				
\usepackage{amsmath, amsgen,amsfonts,amssymb,amsbsy}	

\def\CQG{\it Class.\ Quantum Grav.\ }
\def\PRD{\it Phys.\ Rev.\ D\ }
\def\GRG{\it Gen.\ Rel.\ Grav.\ }
\def\JMP{\it J.\ Math.\ Phys.\ }

\begin{document}

\thispagestyle{plain}		

\title{{\it pp}-waves in conformal Killing gravity}
\author{Alan Barnes\\ 
26 Havannah Lane, \\
Congleton CW12 2EA, \\
United Kingdom. \\
E-Mail: \ \ {\tt Alan.Barnes45678{\bf @}gmail.com}}
\maketitle

\begin{abstract}
\noindent Recently Harada has proposed a gravitational theory which is of third order in
the derivatives of the metric tensor. This has attracted some attention particularly as
it predicts a late-time transition from cosmological decelaration to accelerated expansion
without assuming the presence of dark energy or a non-zero cosmological constant. This
theory has been dubbed \emph{conformal Killing gravity} by Mantica \& Molinari.

The most general exact solutions of the Harada field equations are known for a number
of important physical situations: homogeneous and isotropic cosmological models, static
spherically symmetric vacuum and electrovac spacetimes. These are analogues of the well-known
FRWL, Schwarzschild and Reissner-Nordstr\"om metrics of General Relativity(GR). In this study
the pp-waves in Harada's theory are studied and the most general exact solution is obtained
together with its specialisation for plane waves. The generalisation from GR to Harada's theory
turns out to be straightforward and the solutions only involve an extra non-propagating term.
The solutions have Petrov type N (or 0) and the Ricci tensor is either zero or the Segr\'e type
is [(211)] with zero eigenvalue. 

For any metric in conformal Killing gravity it is shown that more than one possible matter source
can generate the solution. If the metric admits one or more Killing vectors, the ambiguity
in the possible matter sources increases.
\end{abstract}

\vspace{-5 pt}
\section{Introduction}
\noindent Recently Harada\cite{harada} proposed a new gravitational theory which is third-order
in the derivatives of the metric. The field equations of the theory involve the totally symmetric
derivative of a trace-modified  Einstein tensor $\tilde{G}_{ab}$
\begin{equation}
  H_{abc} = \tilde{G}_{(ab;c)}\quad\mathrm{where} \quad
  \tilde{G}_{ab} = R_{ab}-\frac{1}{3}Rg_{ab}=   G_{ab}-\frac{1}{6}Gg_{ab},
  \label{tildeG}
\end{equation}
and of the similarly modified energy-momentum tensor
\begin{equation}
  T_{abc} = \tilde{T}_{(ab;c)} \quad\mathrm{where} \quad
  \tilde{T}_{ab} = T_{ab}-\frac{1}{6}Tg_{ab}
  \label{tildeT}
\end{equation}
where round brackets indicate symmetrisation. The field equations are
\begin{equation}
  H_{abc} = T_{abc}.
\label{hfe}
\end{equation}
The energy-momentum conservation equation follows from \eqref{hfe} by contraction:
\[ g^{ac}H_{abc}=G^a_{b;a} = 0 = g^{ac}T_{abc} =T^a_{b;a}. \]
It also follows immediately  that any solution of the Einstein field equations(EFEs)
$G_{ab}=T_{ab}$ automatically satisfies Harada's field equations\eqref{hfe}. A similar conclusion
holds for solutions of the EFEs with a cosmological constant $G_{ab}+\Lambda g_{ab}=T_{ab}$.
The vacuum solutions are characterised by the condition $T_{ab}=0$ so that
$H_{abc}=T_{abc}=0$.

The theory has attracted considerable interest primarily because Harada\cite{harada}
showed that even in the case of a matter-dominated universe ($p=0$) with $\Lambda=0$
there was a transition from decelarating to accelerating expansion. In a
second paper Harada\cite{harad1} considered this problem in greater depth and
suggested that his theory also had the potential to address the Hubble tension
problem. Mantica \& Molinari\cite{mantmol} examined Harada's field equations and
showed that they can be recast in the form of Einstein's field equations with an
additional source term which is a second order gradient conformal Killing tensor $C_{ab}$
defined by $C_{ab} = G_{ab}-T_{ab}$. They dubbed the theory \emph{conformal
Killing gravity} as a convenient means of distinguishing it from an earlier theory
proposed by Harada\cite{harad2} namely \emph{Cotton gravity}. Mantica \& Molinari used
their reformulation of Harada's theory and their earlier work\cite{mantmol2} on conformal
Killing tensors in Robertson-Walker spacetimes to rederive
Harada's first integral of the evolution equations for the scale factor $a(t)$ and to
independently obtain the results of \cite{harada}.

Harada\cite{harada} derived an analogue of the Schwarzschild
solution in GR and later Tarciso et al.\cite{tarciso} derived an analogue of
the  Reissner-Nordstr\"om solution. Barnes\cite{barnes}\cite{barne1} pointed out that
these solutions were not the most general static vacuum and electrovac
spherically symmetric solutions and derived the general solutions in both cases
although the solutions involved two power series.
Barnes\cite{barne2} also proved the analogue of Birkhoff's theorem for a number of special cases 
of non-static spherically symmetric vacuum metrics, but was unable to prove the general theorem.
Very recently Clement \& Nouicer\cite{c-n} found closed form solutions for the vacuum and electrovac
metrics and in the same paper exhibited counter-examples to Birkhoff's theorem. These are non-static
vacuum FRWL metrics. 

One of the aims of this study is to derive exact solutions for {\it pp}-waves and \emph{a fortiori}
plane waves in conformal Killing gravity(CKG). These are generally interpreted to represent
gravitational waves far from their source.
Secondly it is shown that the matter source for any metric in conformal Killing gravity is only
defined  up to the addition of a cosmolgical constant (or dark energy) term. Furthermore if the
metric admits one or more Killing vectors or second order Killing tensors there are
correspondingly more possible matter sources for the metric.

\section{Plane-fronted waves with parallel rays}
\noindent The plane-fronted waves with parallel rays ({\it pp}-waves) studied in this paper are spacetimes admitting a covariantly constant null bivector field $W_{ab}$:
\begin{equation}
 W_{ab;c}=0,  \mathrm{\ with\ } W_{ab}= p_{[a}k_{b]},\  k_ak^a=p_ak^a=0 \mathrm{\ and\ }
p_ap^a=-1.
\label{w-eq}
\end{equation}
This definition\eqref{w-eq} implies $k^a$ is a covariantly constant null vector, but the converse is not true. Thus the waves studied here are a subclass of general  {\it pp}-waves.

 In any spacetime admitting a covariantly-constant null bivector using complex null coordinates the metric imay be written as\cite{e-k}
\begin{equation}
 \mathrm{d}s^2 = 2\mathrm{d}u\mathrm{d}v+2H\mathrm{d}u^2-\mathrm{d}z\mathrm{d}\bar z,
  \label{metric}
\end{equation}
where $H=H(z,\bar z, u)$. 
They were first studied by Brinkmann\cite{brink} and have been studied extensively since then.  There are thorough reviews of {\it pp}-waves in GR by Ehlers \& Kundt\cite{e-k} and by
Stephani et al.\cite{exact-sol} which are quoted extensively below.

Using the complex-null tetrad $({\bf l},{\bf n}, {\bf m},{\bf \bar m})$
of one forms:
\begin{equation}
  {\bf l}=H \mathrm{d}u +\mathrm{d}v,\qquad {\bf k}=\mathrm{d}u,
 \qquad {\bf m} = \mathrm{d} z, \qquad{\bf \bar m}=\mathrm{d}\bar z.
\label{tetrad}
\end{equation}
the only non-zero Newman-Penrose\cite{np} component of the trace-free Ricci tensor is
\begin{equation}
  \Phi_{22'} = R_{11}/2 = 2H_{z\bar z}.
  \label{ric}
\end{equation}
The Ricci scalar $R$ is zero and the only non-zero Newman-Penrose component of the Weyl tensor is
\begin{equation}
  \Psi_4 = 2H_{\bar z\bar z}.
  \label{psi}
\end{equation}
The Petrov type is N unless $H$ is linear in $z$ and $\bar z$ when the metric is conformally flat.
Plane waves are defined to be {\it pp}-waves for which $\Psi_{4,\bar z}=0$ and $\Phi_{22',\bar z}=0$.

The form of the metric\eqref{metric} is preserved under the coordinate
transformations\cite{exact-sol}:
\begin{eqnarray}
z'= \exp(\mathrm{i}\alpha)(z+2\beta(u)),
  \quad& v'=a\left(v+\dot \beta(u)\bar z+\dot{\bar \beta}(u)z+\gamma(u)\right),\nonumber\\
u' = (u+u_0)/a,\quad&
H' = a^2\left(H-\ddot\beta(u)\bar z-\ddot{\bar\beta}(u)z+2\dot\beta\dot{\bar\beta}-\dot \gamma(u)\right),
  \label{coord-trans}
\end{eqnarray}
where $\alpha$, $a$ and $u_0$ are real constants, $\gamma(u)$ is real and $\beta(u)$ is complex.

Any {\it pp}-wave admits the null Killing vector ${\bf k} = \partial_v$, but for special forms of
$H(u,z,\bar z)$ there may be up to 6 additional Killing vectors.
The symmetry classes {\it pp}-waves were thoroughly investigated by Sippel \& Goenner\cite{sip-goen}
who generalised earlier work by Ehlers \& Kundt\cite{e-k} for the vacuum case in GR.

\subsection{{\it pp}-waves in GR}
\noindent The only possible matter sources are vacuum $T_{ab}=0$ or pure radiation
$T_{ab} =B(u,z,\bar z)k_ak_b$ with $B>0$. Note that $T^a_{b;a}=0$ implies $B$ is independent of $v$.
Pure radiation fields include null Einstein-Maxwell fields with vector potential
$A_a=A(u,z,\bar z)u_{,a}$. Maxwell equations imply that $A_{,z\bar z}=0$.
Hence $A(u,z,\bar z)$ is an arbitrary $u$-dependent harmonic function of $z$ and $\bar z$. Thus 
 $A = F(z,u)+\bar{F}(\bar z,u)$, where $F$ is an arbitrary function analytic in $z$.
The energy-momentum tensor is given by
\begin{equation}
  T_{ab}=8A_{,z}A_{,\bar z}.
  \label{Tem}
  \end{equation}
For the vacuum case it follows from \eqref{ric} that $H$ is a general $u$-dependent harmonic
function
\begin{equation}
  H = f(z,u)+\bar{f}(\bar z, u),
  \label{gr-vac}
\end{equation}
where  $f$ is an arbitrary function analytic in $z$. Similarly, if the source is a null
electromagnetic field, it follows from \eqref{ric} \& \eqref{Tem} that
\begin{equation}
  H = f(z,u)+\bar{f}(\bar z, u) + F(z,u)\bar{F}(\bar z,u).
  \label{gr-em}
\end{equation}
For a general pure radiation field $H_{,z\bar z}=B$. So essentially $H$  may be an arbitrary
function of $u$, $z$ and $\bar z$.

For  plane Einstein-Maxwell waves $\Psi_{4,\bar z}=0$ and $\Phi_{22',\bar z} =0$ and so
\eqref{ric} and \eqref{psi} imply that $H_{,\bar z\bar z z} = H_{,\bar z\bar z\bar z}=0$. Thus $H$ is
at most quadratic in $z$ and $\bar z$. Except in the conformally flat case where $H$ is linear in
$z$ and $\bar z$, the coordinate freedom \eqref{coord-trans} can be used to simplify $H$ to the
form:
\begin{equation}
  H = C(u)z^2+\bar{C}(u)\bar z^2+D(u)z\bar z,
\end{equation}  
where $C$ and $D$ are complex and real functions of $u$ respectively and where $D=0$ in the vacuum
case. 

\subsection{{\it pp}-waves in conformal Killing gravity}
\noindent For the metric \eqref{metric} the only \emph{frame} components of $H_{abc}$
that are not identically zero are
\begin{equation}
  H_{111} = 3H_{,z\bar z u}\quad H_{112}  =  H_{121} = H_{211}=H_{,z\bar z z}
  \quad H_{113}  =  H_{131} = H_{311} = H_{,z\bar z\bar z}.
 \end{equation}
Thus for vacuum {\it pp}-wave solutions of \eqref{hfe}, $H_{,z\bar z}=c$ where $c$ is a constant
which is non-zero for solutions which do not also satisfy the EFEs. From \eqref{ric} it follows
that the Ricci tensor is given by $R_{ab} = ck_ak_b$ which in GR would be interpreted as a
covariantly-constant pure radiation field.
The general form of $H$ is
\begin{equation}
  H =  f(z,u)+\bar{f}(\bar z, u)+c z\bar z.
  \label{hara1}
\end{equation}
Specialising to  type N plane waves
\begin{equation}
    H = C(u)z^2+\bar C(u)\bar z^2 +c z\bar z,
    \label{hara2}
\end{equation}
where linear terms have been set to zero by means of the coordinate
transformations\eqref{coord-trans}.

Putting $z=x+\mathrm{i}y$ in \eqref{hara2} one obtains
\begin{equation}
  H = g(u)(x^2-y^2) + h(u)x y +c(x^2+y^2),
\end{equation}
where $g$ and $h$ are arbitrary real functions of $u$. There is also a
non-propagating term quadratic in $x$ and $y$. For {\it pp}-waves the results are similar with
only a non-propagating quadratic term in addition to the GR form of $H$.

For pure radiation fields $T_{ab}=Bk_ak_b$ the only components of $T_{abc}$ which are not identically
zero are
\begin{equation}
  T_{111}=B_{,u},\quad T_{112}=T_{121}=T_{211} = B_{,z},\quad T_{113}=T_{131}=T_{311} = B_{,\bar z}.
\end{equation}
From the field equations\eqref{hfe} it follows that $H_{,z\bar z}=B+c$. Thus {\it pp}-wave solutions
in CKG differ from those in GR by only an additional term $cz\bar z$ in $H$.

\subsection{The conformally flat case}
\noindent From \eqref{psi} the metric\eqref{metric} is conformally flat if $H_{,\bar z\bar z}=0$.
Hence, as $H$ is real,  $H_{,zz}=0$. Thus
\begin{equation}
  H = \delta(u)z\bar z+ \eta(u)z+\bar\eta(u)\bar z+\phi(u),
  \label{cfcase}
\end{equation}
where $\eta$ is complex and $\phi$ \& $\delta$ are real. From \eqref{ric}
$R_{11} = 4\delta(u)$. If $\delta=0$, the metric is flat. Otherwise as $R_{11}$ is independent of
$z$ and $\bar z$, the solution is a plane wave and, using the coordinate transformations
\eqref{coord-trans}, $H$ may be simplified to $H=\delta(u)z\bar z$.

In GR the only possible matter source is a pure radiation field whose amplitude is independent of
the spatial coordinates. However, in CKG there are non-flat vacuum solutions with $\delta(u)=c$
($c$ constant) in \eqref{cfcase}. Finally the similarity between the solutions in GR and CKG means
that the results of Sippel \& Goenner\cite{sip-goen} on the symmetry classes in GR may be applied
directly in CKG. 

\section{Killing Vectors in Conformal Killing Gravity}
\noindent Barnes\cite{barnes} pointed out that all metrics in CKG which
are not solutions of the GR field equations (with or without a cosmological constant) admit
a non-trivial Killing tensor $\tilde G_{ab}-\tilde T_{ab}$ where $\tilde G_{ab}$ and $\tilde T_{ab}$
are the trace-modified tensors defined in \eqref{tildeG} \& \eqref{tildeT}.
vector
The existence of Killing vectors and tensors has further implications for CKG. If $\tilde T_{ab}$
and $\tilde T'_{ab}$ are trace-modified energy-momentum tensors such that
$\tilde T_{ab}-\tilde T'_{ab}=K_{ab}$ is a Killing tensor, the tensor $H_{abc}$ in the field
equations\eqref{hfe} is the same whether $T_{ab}$ or $T'_{ab}$ is regarded as the matter source.
Note that the trace-modified tensor $\tilde T_{ab}$ uniquely determines $T_{ab}$ since
$T_{ab} = \tilde T_{ab} +1/2\tilde Tg_{ab}$.  To summarise the \emph{same metric may result from
physically distinct matter sources}.

Mantica \& Molinari's observation\cite{mantmol} that any solution in CKG can also be regarded as a
solution in GR with an extra source term which is a conformal Killing tensor (CKT) is one
manifestation of this.
For any metric in CKG with matter source $T_{ab}$, $K_{ab}=\tilde G_{ab}-\tilde T_{ab}$ is a
Killing tensor. Thus $G_{ab}-T_{ab}=K_{ab}+1/2Kg_{ab}$ and the GR field equations are satisfied with the
extra source term which is the conformal Killing tensor $C_{ab}$ where $C_{ab}= K_{ab}+1/2Kg_{ab}$.
Another consequence is that dark energy with energy-momentum tensor $T_{ab}=\lambda g_{ab}$ where
$\lambda$ is \emph{any} constant produces exactly the same gravitational field as in the vacuum
case $T_{ab}=0$. Thus in CKG one might say \emph{dark energy does not gravitate}. Note there is no
relationship between $\lambda$ and the cosmological constant term that appears as an integration
constant in many exact solutions in CKG.

Whenever the metric admits symmetries generated by Killing vectors $\xi^a_I, I = 1 \ldots N$,
any linear combination of these (and the metric) with constant coefficients is a reducible Killing
tensor\cite{Rani} $K_{ab}$ where
\begin{equation}
  K_{ab} = k g_{ab} +\sum_{I=1}^N\sum_{J=1}^N k_{IJ}\xi_{I(a}\xi_{|J|b)},
  \label{red-kt}
\end{equation}
where $k$ and the $k_{IJ}$ are all constants. If there are $N$ Killing vectors, there are 
$1+(N+1)N/2$ possible terms in \eqref{red-kt} which may contribute to the matter source
$\tilde{T}_{ab}$. Some metrics such as the Kerr metric admit one or more irreducible
Killing tensors and there are then corresponding more potential matter sources.

All known exact solutions in CKG suffer from this ambiguity as they all admit one or more Killing
vectors.  For example {\it pp}-wave metrics admit a Killing vector $k^a$. Thus the tensor
\begin{equation}
  T_{ab} = \Lambda g_{ab} + B k_ak_b\qquad\mathrm{where\ } \Lambda \mathrm{\ and\ }B
  \mathrm{\ are\ constants}
  \label{alt-s1}
\end{equation}
satisfies the equation $T_{abc}=0$ and so can be regarded as a matter source for them.
Equation\eqref{alt-s1} describes a dark energy term plus a pure radiation field of constant
amplitude $B$. For plane waves the order of the isometry group is at least 5 and hence they
have many possible matter sources.

Static spherically symmetric vacuum solutions\cite{harada,barnes} admit the Killing vector
$\xi^a=\delta^a_4$. Thus $\tilde T_{ab} = k\xi_a\xi_b$ is a Killing tensor. If $g_{tt} = V(r)^2$ and
$u^a$ is the unit timelike vector parallel to $\xi^a$, the energy-momentum tensor
\begin{equation}
  T_{ab} = kV^2u_au_b +kV^2g_{ab}/2 +\Lambda g_{ab}vector
  \label{alt-s2}
\end{equation}
may be regarded as an alternative matter source for the vacuum solution. It corresponds
to a perfect fluid with energy-density and pressure given by
\begin{equation}
  \rho = \frac{3kV^2}{2}, \qquad p = -\frac{kV^2}{2},
\end{equation}
respectively, plus an arbitrary dark energy term $T_{ab}=\Lambda g_{ab}$.
Static and spherically symmetric metrics actually admit four Killing vectors. Thus there are a total
of 11 possible terms in the Killing tensor \eqref{red-kt}. Consequently there are up to eleven possible
terms contributing to energy-momentum tensors which are possible alternative matter sources for the
vacuum solution. However, only three of these inherit the spherical symmetry of the metric. These are
those already given in \eqref{alt-s2} plus one corresponding to the Killing tensor/trace modified
energy-momentum tensor:
\begin{equation}
  \tilde T_{ab} = c(\xi^1_a\xi^1_b+\xi^2_a\xi^2_b+\xi^3_a\xi^3_b)
     = cr^4(\delta^\theta_a\delta^\theta_b+\sin^2\theta\delta^\phi_a\delta^\phi_b),
\end{equation}
where $\xi^N_a, N=1\ldots 3$ are the three rotational Killing vectors and $c$ is an arbitrary constant.

If there are no symmetries the only ambiguity regarding the matter source $T_{ab}$ of the
metric is an ubiquitous dark energy term of the form $\lambda g_{ab}$.

\section{Conclusions}
\noindent The most general exact solutions for {\it pp}-waves and plane waves
in conformal Killing gravity are obtained. They are straightforward generalisations of
the corresponding solutions in General Relativity with only an additional non-propagating
term quadratic in the spatial coordinates $x$ and $y$.

For all metrics in conformal Killing gravity there is necessarily an ambiguity regarding the matter
source. For metrics without symmetries the matter source is only
determined up to the addition of a cosmological constant (or dark energy) term. However, for metrics
admitting Killing vectors or second-order Killing tensors there are several possible matter sources. 

\section*{Acknowledgements}
\noindent The calculations in {\S}2 were performed using the Sheep/Classi package\cite{Aman} 
which was kindly supplied to me by Jan {\AA}man. I would also like to thank him for
useful discussions on some undocumented features of the system. Thanks also go to Graham Hall of
Aberdeen University for drawing my attention to the work of Goenner and Sippel.

\end{document}